\pdfoutput=1
\documentclass[conference,a4paper]{IEEEtran}
\usepackage[T1]{fontenc}
\usepackage[utf8]{inputenc}
\usepackage{amsmath,amssymb,graphicx,cite}
\usepackage[hidelinks]{hyperref}
\title{AI for Science with GPT-6 Astra: Thermal Design and Electrothermal Analysis of 2D CFET}
\hypersetup{pdftitle={AI for Science with GPT-6 Astra: Thermal Design and Electrothermal Analysis of 2D CFET}}
\author{
\IEEEauthorblockN{Min-Hui Kim\textsuperscript{1}, Khushi Sharma\textsuperscript{2}, Sarah Zhang\textsuperscript{3},
Ye Wang\textsuperscript{4}\textsuperscript{*}}
\IEEEauthorblockA{\textsuperscript{1}Graduate School of Semiconductor Materials and Devices,\\
Ulsan National Institute of Science and Technology (UNIST), Ulsan 44919, Republic of Korea\\
\textsuperscript{2}Materials Science and Engineering, National University of Singapore, Singapore\\
\textsuperscript{3}Materials Science and Engineering, Cornell University, Ithaca, NY, USA\\
\textsuperscript{4}Department of Applied Physics and Science Education,\\
Technische Universiteit Eindhoven, Eindhoven, The Netherlands\\
\textsuperscript{*}Corresponding author: y.wang19@tue.nl\\
The first three authors are listed alphabetically by surname.}
}
\begin{document}
\maketitle
\begin{center}
\footnotesize\emph{This work has been submitted to the IEEE for possible publication. Copyright may be transferred without notice, after which this version may no longer be accessible.}
\end{center}
\begin{abstract}
Thermal optimization of 2D CFET inverters requires testing structural proposals against their electrical costs. We examine these research tasks using an AI agent workflow within a supplied electrothermal model. At 12 nm, Astra selects a redistributed source-interconnect geometry, while a coordinating agent proposes a substrate-directed heat-removal path. The combined design reduces peak temperature rise by 1.67 K at fixed metal volume and 20 $\mu$W. A subsequent metal-resistance sensitivity gives about 0.6-K inverter cooling alongside a 2\% nFET on-current loss. Effective contact-length scaling further shows that lower temperature can accompany higher thermal resistance when current falls. Reproduction identifies agreeing implementations and retains a 104.95-K failure for diagnosis. These results show that an AI scientist workflow can propose thermal structures, test them under common constraints, and quantify their electrical cost.
\end{abstract}
\begin{IEEEkeywords}
GPT-6, AI for science, two-dimensional materials, 2D CFET, electrothermal modeling, thermal management.
\end{IEEEkeywords}

\section{Introduction}
Two-dimensional channels support strong electrostatic control and dense complementary field-effect transistor (CFET) integration~\cite{cfet}. Stacking n- and p-channel devices concentrates heat within one footprint, where dielectrics and thermal boundary resistances hinder its removal~\cite{network,multitier,yalon}. In an inverter, temperature also changes transistor currents and redistributes power. Thermal designs must therefore be evaluated under circuit operation as well as a prescribed heat load.

Finding that design requires deciding which geometry to test, how to compare candidates fairly, and what to investigate next. Researchers ordinarily make these decisions when using TCAD solvers~\cite{tcad}. AI-scientist frameworks that connect hypothesis generation to code execution~\cite{aiscientist} motivate testing whether agents can undertake such tasks in CMOS logic research.

Within a supplied Python model and fixed metal budget, Astra searches interconnect geometries and a coordinating agent proposes a via extension. We evaluate the resulting inverter, distinguish the agents' contributions through search records, and check implementation consistency through reproduction.

\begin{figure}[!t]
\centering\includegraphics[width=\columnwidth]{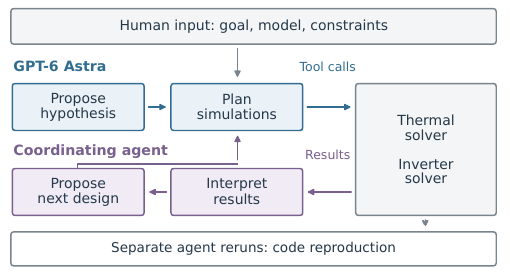}
\caption{Agent workflow within a supplied model. Astra plans the interconnect search; the coordinating agent interprets the trend and proposes the via extension. Tool calls execute the tests; separate reruns check code consistency.}
\label{fig:ai}
\end{figure}

\begin{figure*}[t]
\centering\includegraphics[width=\textwidth]{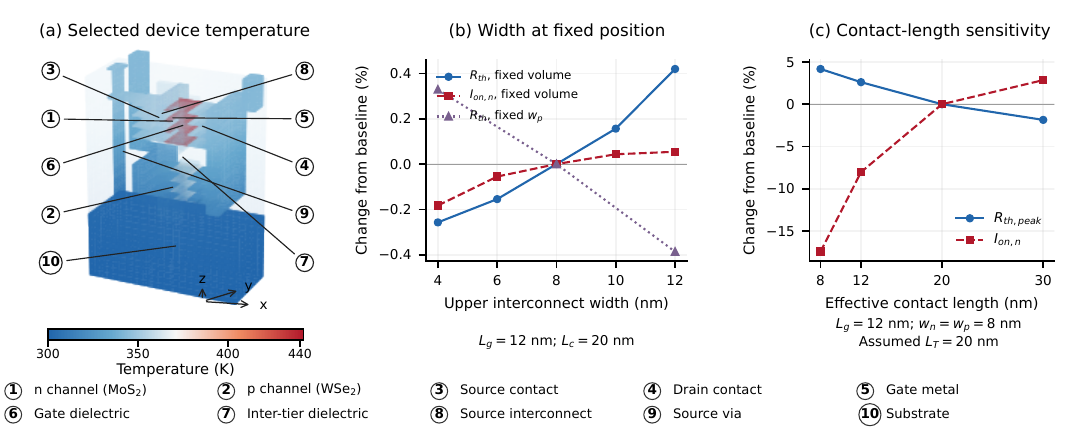}
\caption{Thermal design and electrical cost. (a) Selected device at 20 $\mu$W, $P_n/P_p=2$, $R_{cm}=5$ m$^2$K/GW. At $L_g=12$ nm, (b) upper width varies at fixed metal volume or fixed lower width; (c) effective contact length varies. Changes are relative to $w_n=w_p=8$ nm, $L_c=20$ nm. $R_{th,peak}$ uses fixed 20 $\mu$W; self-heated $I_{on,n}$ includes metal resistance at $V_{GS}=V_{DS}=0.7$ V.}
\label{fig:thermal}
\end{figure*}

\begin{figure*}[t]
\centering\includegraphics[width=\textwidth]{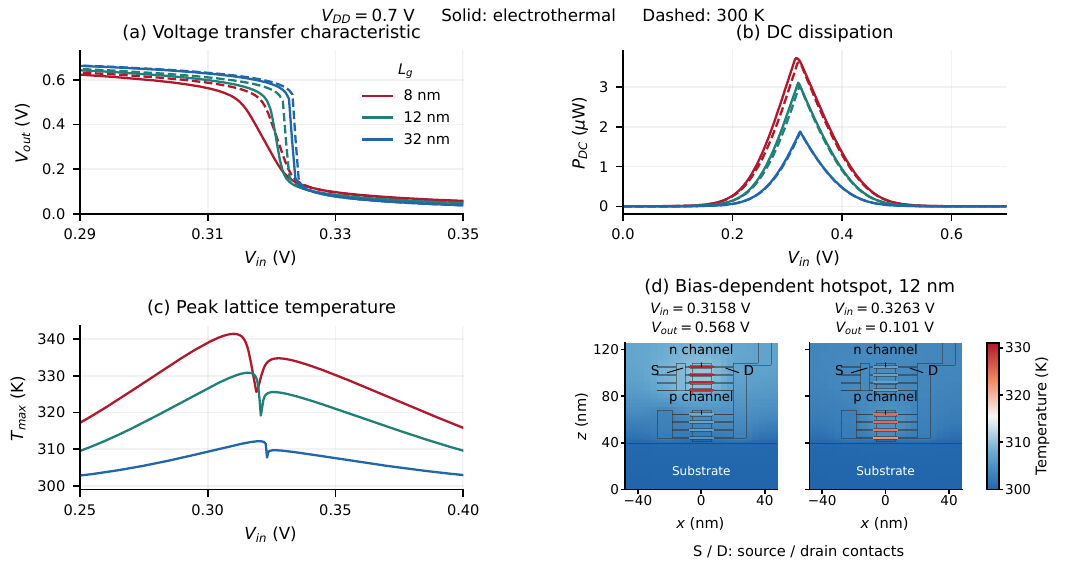}
\caption{DC inverter response without metal resistance at $V_{DD}=0.7$ V. (a) Voltage transfer characteristics, (b) DC power and (c) peak lattice temperature for baseline interconnects. Colors denote gate length; solid curves include self-heating, dashed curves are isothermal at 300 K. (d) Selected 12-nm device sections at $y=0$ under two static biases, with a common temperature scale and emphasized channels.}
\label{fig:scaling}
\end{figure*}

\section{Methodology and Device Model}
\subsection{Thermal Model and Electrical Loading}
The search uses four MoS$_2$ sheets above four WSe$_2$ sheets ($L_g=12$ nm, width 40 nm, heated length 16 nm; Table~\ref{tab:device_compare}). Source local interconnects join each tier's sheet contacts; their widths vary while contact length stays fixed. Channel cross-plane, metal and dielectric conductivities are 2, 25 and 1.2 W\,m$^{-1}$K$^{-1}$.

We solve anisotropic steady-state heat conduction by finite volumes, including thermal boundary resistance. The bottom is fixed at 300 K; the top exchanges heat with a 300-K bath (Table~\ref{tab:device_compare}). Fixed-power tests use 20 $\mu$W, 30\% contact dissipation and $R_{cm}=5$ m$^2$K/GW, with constant conductivities and no interconnect Joule heating. Energy balance, source normalization and reciprocity are checked. Refining the original fixed-power comparison changes its improvement by 0.004 K; monolayers remain one cell thick.

To test circuit operation, we instead determine power from the inverter currents at $V_{DD}=0.7$ V. Gate lengths are 8, 12 and 32 nm; follow-up width/contact scans cover all three. Both gates share an input, and their interconnected drains form the output. We solve
\begin{equation}
I_n(V_{in},V_{out},T_n)=I_p(V_{DD}-V_{in},V_{DD}-V_{out},T_p)
\end{equation}
together with $\mathbf T=300~\mathrm K+\mathbf H\mathbf p$, where $\mathbf H$ maps channel/contact powers to channel-average temperatures. The charge-sheet current model accounts for finite oxide thickness~\cite{frank,gilardi}, mobility proportional to $T^{-1.35}$ and saturation velocity proportional to $T^{-0.5}$. The n/p width-normalized contact resistances are 500/800 $\Omega\,\mu$m per contact. Their voltage drops determine contact heating, replacing the prescribed 30\% fraction; current balance gives total power $P_{DC}=V_{DD}I$.

A follow-up sensitivity adds finite-volume metal conduction, voltage drops and Joule heating, assuming $\rho_m=2.93\times10^{-7}$ $\Omega$m. Effective contact-length scaling uses $R_c(L_c)/R_c(L_0)=\tanh(L_0/L_T)/\tanh(L_c/L_T)$~\cite{contactlength}, with $L_0=L_T=20$ nm and the above reference resistances. Injection beneath a top contact is not resolved. Width scans compare fixed metal volume with fixed lower width; contact-length scans change footprint and metal volume.

\subsection{Agent Roles and Structural Search}
Astra tests whether widening the hotter upper source interconnect and moving it inward improves cooling, using the requested Ultra setting (Fig.~\ref{fig:ai}). The objective $J(d)=\max_{r\in\{0.5,1,2\}}\Delta T_{\max}(d,r)$ minimizes the worst peak rise above 300 K over power ratios $r=P_n/P_p$. All candidates preserve the same metal volume, top metal area, heat sources and thermal boundaries.

The coordinating agent identifies the reversed width trend and proposes substrate-directed vias. No equivalent Astra acknowledgement is documented, so its sweep is distinguished from this interpretation. Design families were specified before evaluation and candidate results retained; the search was adaptive, not blind.

\section{Results and Discussion}
\subsection{Source Interconnects and Heat Flow}
\begin{table}[!t]
\caption{Thermal performance of design stages on a common mesh}
\label{tab:prediction}
\centering\small
\setlength{\tabcolsep}{4pt}
\begin{tabular}{@{}lrr@{}}
\hline
Structure & $\Delta T_{\max}$ (K) & Reduction (K) \\
\hline
Baseline & 141.63 & 0.00 \\
Redistributed interconnect & 140.75 & 0.88 \\
Substrate-directed path & 140.83 & 0.80 \\
Combined & 139.96 & 1.67 \\
\hline
\end{tabular}
\par\smallskip\raggedright\footnotesize
n-on-p; $P=20$ $\mu$W, $R_{cm}=5$ m$^2$K/GW; 43,493 cells. Metal volume, top area, heat sources and boundary conditions are fixed. All four peak at $r=2$ over $r=0.5,1,2$.
\end{table}

The selected design narrows and repositions the upper source interconnect while reallocating the saved metal to the lower tier. Peak rise falls by 0.88 K (Table~\ref{tab:prediction}), contradicting the initial widening hypothesis. Width and position change together, while semiconductor contacts remain fixed.

To improve downward heat flow, the coordinating agent adds a narrow via directed toward the substrate [Fig.~\ref{fig:thermal}(a)]. With the interconnect change and the same metal budget, peak rise falls from 141.63 to 139.96 K. The 1.67-K (1.18\%) reduction exceeds either separate change and persists under refinement.

The fixed-position control distinguishes narrowing from redistribution [Fig.~\ref{fig:thermal}(b)]. Changing an 8/8-nm upper/lower pair to 4/12 nm reduces thermal resistance by 0.26\%, but narrowing only the upper interconnect raises it by 0.33\%. The benefit therefore depends on metal allocation. Contact scaling further separates temperature from performance [Fig.~\ref{fig:thermal}(c)]: shortening $L_c$ from 20 to 8 nm raises thermal resistance by 4.2\%, yet lowers inverter peak temperature by 6.2 K as nFET on-current falls by 17\%. Lower temperature alone is not evidence of better heat removal.

\subsection{CFET Inverter Electrothermal Operation}
The fixed-power result establishes improved heat removal. To test whether it persists in an inverter, we let the transistor currents and temperatures determine the heat load together. At $L_g=8$ nm and $V_{gs}=V_{ds}=0.7$ V, self-heating lowers nFET on-current from 245.5 to 208.6 $\mu$A/$\mu$m (15.0\%); the pFET loses 11.5\% under corresponding bias. Currents use the summed sheet width of $4\times40$ nm. These unequal responses shift the inverter switching voltage ($V_{in}=V_{out}$) from 0.3211 to 0.3190 V and reduce the voltage-gain magnitude there from 83.4 to 52.1 [Fig.~\ref{fig:scaling}(a)].

Near the transition, however, thermal broadening of carrier occupation competes with mobility loss, raising peak DC power from 3.62 to 3.73 $\mu$W at 8 nm [Fig.~\ref{fig:scaling}(b)]. Temperature peaks at a different input voltage: $V_{in}=0.3102$ V, versus 0.3163 V for power. There, $V_{out}=0.561$ V places most of the voltage drop across the upper nFET, whose mean temperature reaches 340.7 K while the pFET remains at 309.1 K. Thus total power alone does not determine the hotspot. Maximum temperature decreases from 341.4 to 312.1 K between 8 and 32 nm [Fig.~\ref{fig:scaling}(c)].

\begin{table}[!t]
\caption{Cross-agent reproduction of the fixed thermal reference cases}
\label{tab:reproduction}
\centering\footnotesize
\setlength{\tabcolsep}{3pt}
\renewcommand{\arraystretch}{1.04}
\begin{tabular}{@{}lrrl@{}}
\hline
Model / effort label & $e_T$ (K) & $e_C$ (pp) & Status \\
\hline
GPT-6 / Light$^a$ & 0 (ref.) & 0 (ref.) & Reference \\
GPT-6 Astra / Ultra$^b$ & 1.34e-12 & 3.20e-13 & Pass \\
\hline
Claude Fable 5.1 / high & 4.26e-13 & 2.42e-13 & Pass \\
Claude Opus 5 / high (A) & 4.98e-10 & 2.88e-10 & Pass \\
Claude Opus 5 (B) & 4.26e-13 & 2.42e-13 & Pass \\
GPT-5.6 Sol / max & 4.98e-10 & 2.88e-10 & Pass \\
GPT-5.6 Terra / max$^\dagger$ & 1.39e-12 & 4.12e-13 & Pass \\
GPT-5 / Codex & 1.08e-12 & 8.74e-13 & Pass \\
Claude Haiku 4.5 & 104.95 & 15.16 & \textbf{Fail} \\
Claude Opus 4.8 / middle$^*$ & 4.26e-13 & 2.42e-13 & Pass \\
Claude Sonnet 5 / middle$^*$ & 1.08e-12 & 8.74e-13 & Pass \\
GPT-5.6 Terra / max$^*$ & 1.08e-12 & 8.74e-13 & Pass \\
GPT-5.6 Luna / max$^*$ & 7.96e-13 & 4.12e-13 & Pass \\
Kimi / basic high$^*$ & 7.96e-13 & 4.12e-13 & Pass \\
Perplexity [Grok 4.6]$^*$ & 1.08e-12 & 8.74e-13 & Pass \\
Grok-labelled table$^\ddagger$ & 4.59e-04 & 1.09e-04 & Table only \\
\hline
\end{tabular}
\par\smallskip\raggedright
$e_T$: maximum peak-rise error (four cases); $e_C$: maximum reduction error (two resistances). Pass requires $e_T\leq0.01$ K, $e_C\leq0.01$ pp, residual/energy error $<10^{-7}$ and rerun agreement. $^a$Reported setting. $^b$Requested setting; reference hidden before freezing. $^*$Unverified archive label. $^\dagger$Prior Sol summary visible. $^\ddagger$No code evidence. A/B: separate submissions. Values include rounding; model/effort comparisons are uncontrolled.
\end{table}

At 12 nm, the combined design lowers maximum inverter temperature from 330.75 to 330.41 K. This smaller 0.335-K reduction reflects the baseline's 2.97-$\mu$W load at its temperature maximum. Rescaling the fixed-power source distribution gives 0.248-K cooling; using the inverter's channel/contact fractions gives 0.329 K. Self-consistent feedback and the shift in peak bias bring this to 0.335 K. Relative to temperature rise, the benefit is 1.09\%, close to the fixed-power 1.18\%.

The hotspot moves between tiers with bias [Fig.~\ref{fig:scaling}(d)]. Including metal resistance gives 0.6-K cooling and 2.2\% lower nFET on-current for the combined 12-nm design. The accompanying width/contact scans show the same trade-off at 8 and 32 nm. Assumed $L_T$ values of 5--50 nm give short-contact current penalties of 2--23\%.

Separate 1--5 nm Green-function calculations~\cite{negf,probe,junction,szabo,stieger}, discussed alongside sub-nanometre gate demonstrations~\cite{desai,wu,perucchini} in the additional analysis, are not coupled to this inverter model.

\subsection{Code Reproduction and Alternative Designs}
Four fixed thermal cases test code consistency (Table~\ref{tab:reproduction}). Passing submissions agree within $5\times10^{-10}$ K; fresh Ultra code agrees within $1.34\times10^{-12}$ K. Haiku-labelled code fails by 104.95 K and 15.16 percentage points owing to indexing and resistance-unit errors. Corrections restore agreement; we retain the original failure. Supplied model/effort labels do not establish a controlled effort comparison.

The separate Fable 5.1 C4 design uses one sheet per tier, doubles interface conductances, introduces AlN, lengthens contacts and adds gate metal (Table~\ref{tab:device_compare}). Corrected, refined mean thermal resistance falls by 45.19\%/40.37\% in the top/bottom tiers with 26.12\% more metal. These material/interface changes extend beyond our search; different loads and metrics prevent ranking the agents.

Cross-implementation checks recover the 1.67255-K reduction with the other study's matrix assembly, despite shared discretization and backend. Neither search has been independently repeated or experimentally validated.

\section{Conclusion}
We used an Astra-based AI scientist workflow to design and evaluate 2D CFET thermal structures. At fixed metal volume ($L_g=12$ nm, $L_c=20$ nm), the best tested design redistributes source-interconnect metal between the tiers and introduces a substrate-directed heat-removal path. It lowers peak temperature by 1.67 K at 20 $\mu$W. Including metal resistance gives 0.6-K inverter cooling with 2.2\% lower nFET on-current, revealing the electrical cost of improved heat removal. The workflow therefore produced a testable structural hypothesis and evaluated its circuit-level trade-off; experimental validation remains necessary.

\begin{table}[!t]
\caption{Device parameters for the two design studies}
\label{tab:device_compare}
\centering\footnotesize
\setlength{\tabcolsep}{3pt}
\renewcommand{\arraystretch}{1.03}
\begin{tabular}{@{}lcc@{}}
\hline
Parameter & GPT study & Claude baseline $\to$ C4 \\
\hline
Sheets per tier (total) & \textbf{4 (8)} & \textbf{1 (2)} \\
$L_g,L_h,W$ (nm) & 12, 16, 40 & 20, 30, 50 \\
$t_n,t_p$ (nm) & 0.65, 0.70 & 0.65, 0.65 \\
Gate dielectric (nm) & 1 & 3 (HfO$_2$) \\
Within-tier gap (nm) & 6 & Not applicable \\
Inter-tier gap (nm) & 20 & 26 \\
Contact length (nm) & 20 & $20\to30$ \\
Substrate (nm) & 40 (effective) & 1000 (Si) \\
Domain (nm$^3$) & $96\!\times\!80\!\times\!125.4$ & $300\!\times\!200\!\times\!1291.3$ \\
Sheet $k_\parallel^{n},k_\parallel^{p}$ & 35, 25 & 35, 10 \\
Substrate $k_\parallel,k_\perp$ & 35, 30 & 120, 120 \\
$R_{cm}^{n},R_{cm}^{p}$ & 5, 5 & $(40,50)\to(20,25)$ \\
Channel--oxide TBR$^{n,p}$ & 71, 71 & $(50,55.56)\to(25,27.78)$ \\
Top $h$ (MW m$^{-2}$K$^{-1}$) & 2 & 0 (adiabatic) \\
\hline
\end{tabular}
\par\smallskip\raggedright
Arrows: baseline to candidate. Entries without arrows stay fixed within each study. $L_h$: channel plus access/spacer length. Gaps: sheet surface to surface. $k$: W/mK; TBR: thermal boundary resistance in m$^2$K/GW. Both use n-on-p stacking, a 300-K bottom and adiabatic sides.
\end{table}

\section*{Acknowledgment}
We acknowledge Vina Faramarzi (ASML Netherlands B.V.), who conceived the research idea together with Ye Wang. Part of this project originated in the Eindhoven Semiconductor Summer School 2026, with participation by Min-Hui Kim, Prasanna Prasad Mahajan, Aleksander Ogonowski, Michael Scholl, Khushi Sharma, Viren Sharma, Sarah Zhang and Yichen Zou. We thank Prof. Shihab Al-Daffaie of TU/e for coordinating the summer school and project work. This work was supported by the Technische Universiteit Eindhoven startup grant RF204648 and the Dutch Research Council (NWO), grant no. EINF-19934.

\end{document}